\documentclass[osajnl,twocolumn,showpacs,superscriptaddress,10pt]{revtex4-1} 
\usepackage{amsmath,amssymb,graphicx,hyperref}
\begin{document}

\title{Measuring Mueller matrix of an arbitrary optical element with a fixed set of polarization optics}

\author{Salla Gangi Reddy} \email{Corresponding author: sgreddy@prl.res.in}\affiliation{Physical Research Laboratory, Navarangpura, Ahmedabad, India - 380009}
\author{Shashi Prabhakar}\affiliation{Physical Research Laboratory, Navarangpura, Ahmedabad, India - 380009}
\author{Aadhi A}    \affiliation{Physical Research Laboratory, Navarangpura, Ahmedabad, India - 380009}
\author{Ashok Kumar}\affiliation{Instituto de Fisica, Universidade de Sao Paulo, Sao Paulo, 66318, Brazil}
\author{Megh Shah}\affiliation{Indian Institute of Technology, Roorkee, India - 247667}
\author{R.P.Singh} \email{rpsingh@prl.res.in}\affiliation{Physical Research Laboratory, Navarangpura, Ahmedabad, India - 380009}
\author{R.Simon} \email{simon@imsc.res.in}\affiliation{The Institute of Mathematical Sciences, Chennai, India - 600113}

\begin{abstract}We have described a novel way to determine the Mueller matrix of any optical element by using projection method. For this purpose, we have used two universal SU(2) gadgets for polarization optics to obtain projection matrix directly from the experiment. Mueller matrix has been determined using the experimentally obtained projection matrix for three known optical elements namely free space, half wave plate and quarter wave plate. Experimental matrices are in good agreement with the corresponding theoretical matrices. The error is minimized as the experimental conditions remains same for all measurements since we have used a fixed set of polarization optics i.e. there is no removal or insertion of an optical component during the experiment.
\end{abstract}


\maketitle 

\section{Introduction}

To characterise the polarization changes in a given light beam due to any optical element, one needs to find its Mueller matrix. From the Mueller matrix, we can obtain the information about the retardation, the diattenuation and the depolarization of light after passing through any optical element \cite{chipman, nghosh}. Determination of Mueller matrix have diverse applications in various fields such as tomography \cite{OCT}, micro-electronics \cite{micro-elec}, geo-science \cite{ocean}, astronomy  \cite{astro}, study of polarization changes in vortex beams \cite{rpsir}, liquid crystals and characterisation of undesired modulation introduced by a spatial light modulator \cite{SLM}.  Along with these, determination of Mueller matrix in optical coherence tomography gives the birefringence and the orientation of optic axis with respect to the incident beam for a given sample \cite{tomography}.  \\

  A three component SU(2) gadget for polarization optics is developed by Simon and Mukunda, can be used to generate any polarization state on the Poincar\'e sphere from a given arbitrary input polarization state \cite{Simon}. This gadget called Simon-Mukunda polarization gadget \cite{bagini} consists of one half wave plate (H) and two quarter wave plates (Q); the polarization states can be obtained by changing the rotation of their fast axes appropriately \cite{Neeti}. One can use SM gadget in different configurations such as Q-H-Q, Q-Q-H and H-Q-Q. However, we have used Q-H-Q configuration only. Recently, the principle of SM gadget for determining the polarization states has been used in variety of applications \cite{lidar, sm, pancha}. In the depolarization studies of light beams due to polar meso-spheric cloud, Lidar can introduce additional polarization changes to the receiving signal. To compensate these changes, Hayman et.al. have used the principle of SM gadget \cite{lidar}. Schilling et. al. presented a theoretical scheme to determine the higher order correlations in a single light beam by introducing proper unitary transformations to the input state with SM gadget \cite{sm}. While measuring the Pancharatnam phase of a polarization state with techniques of interferometry and polarimetry, Loredo et. al. have used SM gadget \cite{pancha}. However, in the present study, we have used it to find out the Mueller matrix of a given optical element. We have used two SM gadgets; one for the generation of input polarization states and another for obtaining the projections of each output state on a given set of four input states. To the best of our knowledge, we are the first to demonstrate the use of this gadget to determine the Mueller matrix.

 In conventional approach, Mueller matrix is determined by measuring the output Stoke's parameters of a given light beam, corresponding to four different input Stoke's parameters. These parameters are found by measuring the intensities with different polarizations (horizontal, vertical, left and right circularly polarized and both diagonal). This method is also known as successive probing method \cite{layden} and it needs 24 intensity measurements with known input Stoke's parameters. In Mueller matrix imaging polarimetry (MMIP), one can obtain Mueller matrix by taking 16, 36 or 49 images \cite{mmip}. In one of the earlier studies, to obtain the Mueller matrix, Azzam presented a technique called photo polarimetry that uses two quarter  wave plates (one in polarizing optics and another in analysing optics) rotating at speeds $\omega$  and $5\omega$ \cite{azaam}.  
Alternatively, the present study provides a novel and simple method to determine the Mueller matrix of an optical element using two SM gadgets. The theory of this method is discussed in section 2 and experimentally demonstrated in section 3.

\section{Theoretical Analysis}

 The Stoke's vector describes the complete polarization state of an electromagnetic wave. When an optical beam with Stoke's vector $(S)$ passes through an optical element, the output Stoke's vector $(\tilde{S})$ can be written as 
\begin{equation} \label{Ss}
  \tilde{S} =  MS
\end{equation} where $S$ and $\tilde{S}$ are four element column vectors and M is a $4 \times 4$ matrix defined as Mueller matrix. We form two matrices $\Omega$ and $\tilde{\Omega}$ by arranging the four input and output Stoke's vectors as four columns 
\begin{equation}
  \Omega=[S_{1}~S_{2}~S_{3}~S_{4}], \hspace{1cm} \tilde{\Omega}=[\tilde{S}_{1}~\tilde{S}_{2}~\tilde{S}_{3}~\tilde{S}_{4}].
\end{equation}
Using Eq. (\ref{Ss}), the relationship between $\Omega$ and $\tilde{\Omega}$ can be written as
\begin{equation}
  \tilde{\Omega} =  M\Omega.
\end{equation}
 
The four input states forming $\Omega$, must be on the surface of the Poincar\'e sphere. The projection of each output state $\tilde{S}_{1}, ~\tilde{S}_{2}, ~\tilde{S}_{3}, ~\tilde{S}_{4}$ on input states $S_{1}, ~S_{2}, ~S_{3}, ~S_{4}$ gives the 16 non-negative real numbers which form the elements of projection matrix $\Lambda$. These elements are given by
\begin{equation}
\label{pro}
\Lambda_{ij} = \frac{1}{2} (S_{j})^T \tilde{S_{i}} = \frac{1}{2}\sum_{\alpha=1}^{4} S_{j}^{\alpha}\tilde{S}_{i}^{\alpha}
\end{equation} 
where $i,j=$ 1 to 4 and $S_i^\alpha$ denotes the $\alpha^{th}$ element of $i^{th}$ Stoke's vector and $\Lambda_{ij}$ is the projection of $\tilde{S_i}$ state on $S_j$ state. Therefore, the projection matrix 
\begin{equation}
\Lambda=\frac{1}{2} \Omega^{T} \tilde{\Omega} = \frac{1}{2} \Omega^{T} M \Omega .
\end{equation}
As a consequence, Mueller matrix $M$ can be written as 
\begin{equation} \label{mueller}
  M=2 (\Omega^T)^{-1} \Lambda \Omega^{-1}.
\end{equation}

To obtain Mueller matrix, one needs to choose four input states and determine the corresponding projection matrix. For the sake of completeness, we have considered three tetrahedrons on the Poincar\'e sphere as three sets of input states. The four vertices of each tetrahedron form a set of input states with a maximum value of determinant for $\Omega$ matrix. We have selected these three tetrahedrons to optimize the errors in Mueller matrices \cite{layden}. The optimal input states $(1, q_{i}, u_{i}, v_{i})^T$ for any polarimeter, should obey the following conditions. 
\begin{equation}
\sum_{\i=1}^{4} q_{i} = \sum_{\i=1}^{4} u_{i} = \sum_{\i=1}^{4} v_{i} = 0;    
\end{equation}
 \begin{equation}        
\sum_{\i=1}^{4} q_{i}  u_{i}  = \sum_{\i=1}^{4} u_{i} v_{i} = \sum_{\i=1}^{4} v_{i} q_{i} = 0 \hspace{0.1cm} \rm{and} \hspace{0.1cm}
\end{equation}
\begin{equation}
\sum_{\i=1}^{4} q_{i}^{2} = \sum_{\i=1}^{4} u_{i}^{2} = \sum_{\i=1}^{4} v_{i}^{2} = \frac{4}{3}  
\end{equation}
By solving these equations, one can get a number of optimal states for polarimetry.

 The Stoke's parameters ($ S^1_i , i = 1-4 $) representing the first tetrahedron are given by         
\begin{center}
 $S^1_{1}$ = $\begin{bmatrix}
       1.000  \hspace{0.1cm}         \\[0.3em]
       1.000           \\[0.3em]
       0.000                     \\[0.3em]
       0.000
     \end{bmatrix}$,      
 $S^1_{2}$ = $\begin{bmatrix}
       1            \\[0.3em]
       -0.333         \\[0.3em]
      0.943         \\[0.3em]
       0          
            \end{bmatrix}$, \vspace{0.2cm}            
                
 $S^1_{3}$ = $\begin{bmatrix}
       1            \\[0.3em]
       -0.333          \\[0.3em]
        -0.472                   \\[0.3em]
       0.816           \end{bmatrix}$, 
 $S^1_{4}$ = $\begin{bmatrix}
       1           \\[0.3em]
       -0.334            \\[0.3em]
       -0.471                      \\[0.3em]
       -0.816        
            \end{bmatrix}$     
\end{center}      
and the corresponding Euler's angles are $( 0,0,0 )$, $( 0,0,\kappa_1 )$, $( 0,-\frac{2\pi}{3},\kappa_1 )$ and $( 0,\frac{2\pi}{3},\kappa_1 )$ respectively where $\kappa_1 = \cos^{-1}\left(-0.333\right) = 109.47^\circ$.
The four vertices of the second tetrahedron are represented by the following Stoke's parameters ($ S^2_i , i = 1-4 $)    
\begin{center}  
$S^2_{1}$ = $\begin{bmatrix}
       1           \\[0.3em]
      0.680           \\[0.3em]
      0.701                      \\[0.3em]
      0.214
     \end{bmatrix}$, 
 $S^2_{2}$ = $\begin{bmatrix}
       1            \\[0.3em]
      0.430         \\[0.3em]
       -0.752           \\[0.3em]
     -0.500       
            \end{bmatrix}$, \vspace{0.2cm}            
            
 $S^2_{3}$ = $\begin{bmatrix}
       1            \\[0.3em]
       -0.730            \\[0.3em]
        0.392                    \\[0.3em]
       -0.560              \end{bmatrix}$, 
 $S^2_{4}$ = $\begin{bmatrix}
       1           \\[0.3em]
     -0.390             \\[0.3em]
     -0.350                       \\[0.3em]
      0.852         
            \end{bmatrix}$         
            \end{center} 
and the corresponding Euler's angles are $(0,-\frac{\pi}{10.59},\kappa_2)$, $(0,-\frac{\pi}{5.35},-1.37\kappa_2)$, $(0,-\frac{\pi}{3.27},2.91\kappa_2)$ and $(0,-\frac{\pi}{1.61},2.40\kappa_2)$ respectively where $\kappa_2 = \cos^{-1}\left(0.68\right) = 47.16^\circ$.                 
                       
Similarly, the four vertices of third tetrahedron are represented by   
\begin{center}       
 $S^3_{1}$ = $\begin{bmatrix}
       1           \\[0.3em]
      0.580           \\[0.3em]
       0.576                      \\[0.3em]
       0.580
     \end{bmatrix}$, 
 $S^3_{2}$ = $\begin{bmatrix}
       1            \\[0.3em]
       -0.580         \\[0.3em]
      -0.580           \\[0.3em]
       0.572          
            \end{bmatrix}$, \vspace{0.2cm}            
                      
 $S^3_{3}$ = $\begin{bmatrix}
       1            \\[0.3em]
     0.580            \\[0.3em]
      -0.576                    \\[0.3em]
    -0.572              \end{bmatrix}$, 
 $S^3_{4}$ = $\begin{bmatrix}
       1           \\[0.3em]
       -0.580          \\[0.3em]
       0.580                      \\[0.3em]
       -0.580         
            \end{bmatrix}$   
            \end{center} 
and the corresponding Euler's angles are $(0,-\frac{\pi}{4},\kappa_3)$, $(0,-\frac{\pi}{1.30},2.30\kappa_3)$, $(0,\frac{\pi}{1.30},\kappa_3)$ and $(0,\frac{\pi}{4},2.30\kappa_3)$ respectively where $\kappa_3 = \cos^{-1}\left(0.58\right) = 54.54^\circ$.

\begin{center}
 \begin{table*}   \label{tab:gan}
  \begin{minipage}[b]{1\linewidth}        \centering
 \renewcommand{\arraystretch}{1.4}
  \begin{tabular}{|c|c|c|c|} \hline
\hspace{0.3cm} State $S^k_i$ \hspace{0.3cm} & \hspace{0.7cm} $\theta^k_{iQ1}$  \hspace{0.7cm}  & \hspace{0.7cm} $\theta^k_{iH1} $   \hspace{0.7cm}    &  \hspace{0.7cm} $\theta^k_{iQ1'}$  \hspace{0.7cm}       \\ \hline	 
 
 $S^{1}_{1}$ & $ \frac{\pi}{4} $     & $ -\frac{\pi}{4} $     & $ \frac{\pi}{4} $ \\ \hline
 
$S^{1}_{2}$ & $\frac{\pi}{4} $     & $ -\frac{\pi}{4}-\frac{\kappa_1}{4}$     & $ \frac{\pi}{4}-\frac{\kappa_1}{2}  $ \\ \hline

$S^{1}_{3}$ & $ \frac{\pi}{4} $ & $ -\frac{\pi}{4}-\frac{\kappa_1}{4}+\frac{\pi}{6} $ & $\frac{\pi}{4}-\frac{\kappa_1}{2}$ \\ \hline

$S^{1}_{4}$ &  $ \frac{\pi}{4} $     & $ -\frac{\pi}{4}-\frac{\kappa_1}{4}-\frac{\pi}{6} $     & $ \frac{\pi}{4}-\frac{\kappa_1}{2} $ \\ \hline

$S^{2}_{1}$ & $\frac{\pi}{4}$&$-\frac{\pi}{4}-\frac{\kappa_2}{4}-\frac{\pi}{42.36}$&$ \frac{\pi}{4}-\frac{\kappa_2}{2}$  \\ \hline

$S^{2}_{2}$ & $ \frac{\pi}{4} $& $ -\frac{\pi}{4}+\frac{1.37\kappa_2}{4}-\frac{\pi}{21.41} $ &$\frac{\pi}{4}+\frac{1.37\kappa_2}{2} $  \\ \hline

$S^{2}_{3}$ & $ \frac{\pi}{4} $     & $ -\frac{\pi}{4}-\frac{2.90 \kappa_2}{4}+\frac{\pi}{13.08} $     & $ \frac{\pi}{4}-\frac{2.90 \kappa_2}{2} $ \\ \hline

 $S^{2}_{4}$ & $ \frac{\pi}{4} $     & $ -\frac{\pi}{4}-\frac{2.40 \kappa_2}{4}-\frac{\pi}{6.41} $     & $ \frac{\pi}{4}-\frac{2.40 \kappa_2}{2}$\\ \hline
 
$S^{3}_{1}$ & $ \frac{\pi}{4} $ & $ -\frac{\pi}{4}-\frac{\pi}{16}-\frac{\kappa_3}{4} $ & $ \frac{\pi}{4}-\frac{\kappa_3}{2} $ \\ \hline

$S^{3}_{2}$ & $ \frac{\pi}{4} $     & $ -\frac{\pi}{4}-\frac{\pi}{5.32}-\frac{2.30\kappa_3}{4}$     & $ \frac{\pi}{4}-\frac{2.30\kappa_3}{2} $ \\ \hline

$S^{3}_{3}$ & $ \frac{\pi}{4} $     & $ -\frac{\pi}{4}+\frac{\pi}{5.32}-\frac{\kappa_3}{4} $     & $ \frac{\pi}{4}-\frac{\kappa_3}{2} $  \\ \hline

$S^{3}_{4}$ & $ \frac{\pi}{4} $     & $ -\frac{\pi}{4}+\frac{\pi}{16}-\frac{2.30\kappa_3}{4} $     & $ \frac{\pi}{4}-\frac{2.30\kappa_3}{2} $   \\ \hline
\end{tabular}                                
\caption{ Angles of rotation for wave plates in SM gadget 1 to generate the four vertices of three tetrahedrons.} 
\end{minipage}
\end{table*}  
\end{center}

The sum of four vertices for these tetrahedrons on the Poincar\'e sphere is origin, the center of mass, and corresponds to the maximally mixed (completely un-polarized) state. Optimal states for polarimeter should have minimum norm and condition number along with the maximum determinant. Here, we have considered Frobenius norms and condition numbers due to their direct dependence of errors in Mueller matrix \cite{layden}. These norms and condition numbers of matrix B are determined by formulae 

\begin{equation}
||B||_F = \sqrt{trace(B^{*} B)}
\end{equation}
\begin{equation}
c_F = \rm{norm}(B) \hspace{0.1cm}\rm{norm}(B^{-1})
\end{equation} 
The Frobenius norms, corresponding condition numbers and determinants for three tetrahedrons are (2.8279, 2.8280, 2.8349), (4.4729, 4.4727, 4.4640) and (3.0755, 3.1218, 3.0763) respectively. To generate four input polarization states ($ S_1, S_2, S_3, S_4$) belonging to a particular tetrahedron, we have used first SM gadget. For example, to generate a particular polarization state ($\xi,\eta,\zeta$), the fast axes' orientation of the wave plates in SM gadget are given by the subscripts of Q, H, Q \cite{Neeti}
  \begin{equation}
 \label{SS1}
u(\xi,\eta,\zeta)=Q_{\frac{\pi}{4}+\frac{\xi}{2}}H_{\frac{-\pi}{4}+\frac{\xi+\eta-\zeta}{4}}Q_{\frac{\pi}{4}-\frac{\zeta}{2}}
\end{equation}
where $\xi,\eta,\zeta$ are Euler's angles corresponding to the polarization state. These angles represent the angle made by the frame of reference ($\xi$) from which observations are made, the phase lag between two orthogonal components ($\eta$) in a given polarization state and the rotation ($\zeta$) of that polarization state from horizontal or vertical direction respectively. The angles of $Q_1$, $H_1$, $Q_1'$ in Table (1) corresponding to $S^k_{i}$ generate the $i^{th}$ state of $k^{th}$ tetrahedron.

To obtain the projections of output states from unknown optical element on a given set of input states, we have used another SM gadget. Let $\Lambda^{k}_{ij}$ represents the projection of $\tilde{S_i}^{k}$ state on $S_j^{k}$ state. $\tilde{S_i}^{k}$ state is the output state, obtained after passing $S_i^{k}$ state through the sample. To generate the required polarization state, we kept the angles in first SM gadget as  $ \theta^k_{iQ_1}$, $ \theta^k_{iH_1}$ and $ \theta^k_{iQ'_1}$. To project the output state on $S_j^{k}$ state, we kept the angles in second SM gadget as $ \theta^k_{jQ_1}+\frac{\pi}{2}$, $ \theta^k_{jH_1}+\frac{\pi}{2}$ and $ \theta^k_{jQ'_1}+\frac{\pi}{2}$; it will generate the inverse of that $S_j^{k}$. 
\begin{equation}
(Q_{\theta})^{-1} = Q_{\frac{\pi}{2}+\theta},   \hspace{0.1cm}
 (H_{\theta})^{-1} = H_{\frac{\pi}{2}+\theta} \hspace{0.1cm} \rm{and} \hspace{0.1cm} (Q'_{\theta})^{-1} = Q'_{\frac{\pi}{2}+\theta} 
\end{equation}
The sixteen projection elements corresponding to different orientations of wave plates form the projection matrix. We have done experiment for three sets of input states forming vertices of tetrahedron on the Poincar\'e sphere and corresponding projection matrices are obtained. The reported Mueller matrix is the average of all three Mueller matrices corresponding to three sets of input states.

\section{Experimental Details}

The experimental set-up for the generation of input polarization states and to obtain projection matrix for calculation of the Mueller matrix is shown in Fig. 1. A He-Ne laser having wavelength 632.8 nm and vertical polarization is used for this study. Laser beam is allowed to pass through a polarizer $P_1$ to increase the ratio of vertical to horizontal polarization of the laser and to set the reference for fast axis of all the polarizing elements. Vertically polarized light is then passed though the first SM gadget ($Q_{1}$ $H_{1}$ $Q_{1}'$) to generate four input states. These four states are further passed though a given sample whose Mueller matrix needs to be determined. The projections of resulting states from the sample on the four input states coming from first SM gadget with respect to initial polarization (set by $ P_1 $) are obtained by using a second SM gadget ($Q_{2}$ $H_{2}$ $Q_{2}'$) and an analyzer ($P_2$) whose transmission axis is parallel to the initial polarizer. To obtain this, the wave plates in second SM gadget are rotated in such a way that it will compensate the polarization changes occurred in the initial state while generating them. The angles are obtained simply by adding $90^\circ$ to the angles used in first SM gadget (Table 1). For example, to get the projections of first output state, we fix the first set of angles ($ \frac{\pi}{4} $, $ -\frac{\pi}{4} $, $ \frac{\pi}{4} $) in first SM gadget and take measurements for four sets of angles obtained by adding 90$^\circ$ to the wave plate orientations corresponding to four polarization states of the tetrahedron with second SM gadget. Thus, we repeat the experiment for four sets of angles corresponding to four input states to obtain 16 non-negative projection elements as discussed in Eq. (\ref{pro}). The same procedure is repeated for other two tetrahedrons also. These elements form a $4 \times 4$ projection matrix which is required to determine the Mueller matrix by using Eq. (\ref{mueller}). The resultant Mueller matrix is the average of three matrices corresponding to input states of the three tetrahedrons. An optical multimeter with an accuracy of $0.2\%$ and resolution of 0.01 $\rm{pW}$ is used to measure the output power. We are using three known optical elements $-$ free space, half wave plate and quarter wave plate as samples. We have used zero order quartz wave plates from Melles Griot for making SM gadgets as well as samples for measuring Mueller matrices. These wave plates are mounted on a motorized computer controlled cylindrical rotation stages with a resolution of $0.04^\circ$. The quoted retardance tolerance of the wave plates is $\lambda/200 \hspace{0.1cm} \rm{to} \hspace{0.1cm} \lambda/500$. To begin with, we have aligned the fast axes of all the wave plates as vertical. To assure that the fast axes are perfectly vertical, we put an analyser with its fast axis perpendicular to the initial polarizer. Each optical element is separately aligned in such a way that we get a minimum power in the power meter. 

\begin{center}
\begin{figure} [h]
\includegraphics[width=3.3in]{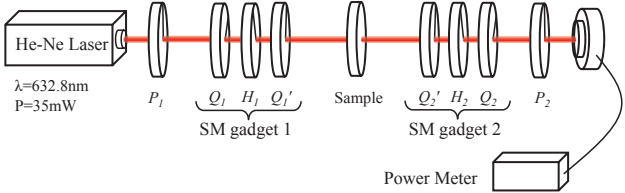}
\caption{ (colour online) Experimental set-up for the determination of Mueller matrix using two SM-gadgets}\label{fig:mueller}
\end{figure} 
\end{center}  

\section{Results and Discussion}

We have experimentally determined the projection matrices for free space, half-wave plate and quarter-wave plate with their fast axes as vertical. The obtained projection matrices, represented by the symbols $ \Lambda^{(iE)}_{free}, \Lambda^{(iE)}_{H}, \Lambda^{(iE)}_{Q}  $ where $i = 1-3$ represents a particular tetrahedron. To obtain Mueller matrices for all three optical elements, we have used projection matrices in Eq. (\ref{mueller}). We have taken the average of three Mueller matrices corresponding to different tetrahedrons for a given sample. The average Mueller matrices from the experiment are denoted with symbols $  M^{(E)}_{free} , M^{(E)}_{H}, M^{(E)}_{Q}  $ for free space ($ free $), half wave plate ($ H $)  and quarter wave plate ($ Q $) respectively.  All the projection and Mueller matrices are normalized with their first element.           

 To obtain the theoretical projection matrices, we have used Jone's matrix analysis. The projection elements represent the fraction of the input intensity of a light beam after passing through all optical elements. They have been obtained from the output Jone's vectors of the beam. The Jone's vector ($J$) of a light beam and corresponding intensity ($I$) is given by    

\begin{center}
 $J = \begin{bmatrix}
       E_{x}        \\[0.3em]
       E_{y}    
 \end{bmatrix},        \hspace{0.5cm}  I =  |E_{x}|^2+|E_{y}|^2 $   
\end{center}

The output Jone's vector has been determined from the product of Jone's matrices of all optical components in the same order as shown in experimental set up (Fig. \ref{fig:mueller}) with input Jone's vector. 
\begin{equation}
J_{output} = J_{P_2} \hspace{0.1cm}   J_{Q_2} \hspace{0.1cm}  J_{H_2} \hspace{0.1cm}  J_{Q_2'} \hspace{0.1cm}  J_{sample} \hspace{0.1cm} J_{Q_1'} \hspace{0.1cm}  J_{H_1} \hspace{0.1cm} J_{Q_1} \hspace{0.1cm} J_{P_1} \hspace{0.1cm} J_{input} 
\end{equation}
We have used the following Jone's matrices for wave plates and polarizers \cite{gold}.
For a wave plate, with retardation $\phi$ and fast axis orientation $\theta$, the Jone's matrix is given by 
\begin{widetext}
\begin{equation}
J_{R}(\phi, \theta) = \begin{bmatrix}
e^{\frac{i \phi}{2}} \cos^2(\theta)+e^{-\frac{i \phi}{2}} \sin^2(\theta)   \hspace{0.2cm} & (e^{\frac{i \phi}{2}}-e^{-\frac{i \phi}{2}}) \cos(\theta) \sin(\theta)   \\   
(e^{\frac{i \phi}{2}}-e^{-\frac{i \phi}{2}}) \cos(\theta) \sin(\theta) &   \hspace{0.2cm} e^{\frac{i \phi}{2}} \sin^2(\theta)+e^{-\frac{i \phi}{2}} \cos^2(\theta) 
\end{bmatrix}
\end{equation}
\end{widetext}
and for a polarizer, with fast axis orientation $\theta$, it is 

\begin{equation}
J_{p}(\theta) = \begin{bmatrix}
 \cos^2(\theta)  \hspace{0.2cm} &  \cos(\theta) \sin(\theta)   \\   
 \cos(\theta) \sin(\theta) &   \hspace{0.2cm} \sin^2(\theta) 
\end{bmatrix}
\end{equation}

The sixteen non-negative and real projection elements have been obtained for 16 sets of orientations of wave plates in SM gadgets which form a projection matrix. The theoretical projection matrices are denoted with symbols $ \Lambda^{(iT)}_{free}, \Lambda^{(iT)}_{H}, \Lambda^{(iT)}_{Q} $ where $i = 1-3$ corresponding to three tetrahedrons. The obtained theoretical projection matrices from output Jone's vectors are used in Eq. (\ref{mueller}) to get the Mueller matrix. The theoretical Mueller matrices have been denoted with symbols $  M^{(T)}_{free}, M^{(T)}_{H}, M^{(T)}_{Q} $. The experimental projection and Mueller matrices are in good agreement with the theoretical matrices shown below. 
\begin{eqnarray}
 \Lambda^{(1E)}_{free} &=& \left[ \begin{array}{r r r r}
       1.000 \hspace{0.3cm}&0.353 \hspace{0.3cm}& 0.341 \hspace{0.3cm}& 0.334 \\[0.3em]
       0.335 \hspace{0.3cm}&1.012 \hspace{0.3cm}& 0.348 \hspace{0.3cm}&0.343 \\[0.3em]
       0.338 \hspace{0.3cm}&0.335 \hspace{0.3cm}&1.003 \hspace{0.3cm}&0.335 \\[0.3em]
       0.332 \hspace{0.3cm}&0.346 \hspace{0.3cm}&0.332 \hspace{0.3cm}&1.003
     \end{array}\right] \nonumber \\               
 \Lambda^{(1E)}_{H} &=& \left[ \begin{array}{r r r r}
       1.000 \hspace{0.3cm}& 0.329  \hspace{0.3cm}&0.322 \hspace{0.3cm}& 0.333        \\[0.3em]
      0.320  \hspace{0.3cm}&0.113  \hspace{0.3cm}&0.784 \hspace{0.3cm}& 0.789      \\[0.3em]
       0.320 \hspace{0.3cm}&0.787 \hspace{0.3cm}& 0.122 \hspace{0.3cm}& 0.790        \\[0.3em]
       0.324 \hspace{0.3cm}&0.782 \hspace{0.3cm}&0.803 \hspace{0.3cm}& 0.123          
     \end{array}\right] \nonumber \\
 \Lambda^{(1E)}_{Q} &=& \left[ \begin{array}{r r r r}
       1.000 \hspace{0.3cm}&0.344 \hspace{0.3cm}&0.306 \hspace{0.3cm}&0.326         \\[0.3em]
       0.350 \hspace{0.3cm}&0.554 \hspace{0.3cm}&0.954 \hspace{0.3cm}&0.171        \\[0.3em]
       0.325 \hspace{0.3cm}&0.168 \hspace{0.3cm}&0.567 \hspace{0.3cm}&0.945         \\[0.3em]
       0.320 \hspace{0.3cm}&0.958 \hspace{0.3cm}&0.210 \hspace{0.3cm}&0.567
     \end{array}\right] \nonumber \\
 \Lambda^{(1T)}_{free} &=& \left[ \begin{array}{r r r r}
       1.000 \hspace{0.3cm}&0.333 \hspace{0.3cm}&0.333 \hspace{0.3cm}&0.333         \\[0.3em]
       0.333 \hspace{0.3cm}&1.000 \hspace{0.3cm}&0.333 \hspace{0.3cm}&0.333         \\[0.3em]
       0.333 \hspace{0.3cm}&0.333 \hspace{0.3cm}&1.000 \hspace{0.3cm}&0.333         \\[0.3em]
       0.333 \hspace{0.3cm}&0.333 \hspace{0.3cm}&0.333 \hspace{0.3cm}&1.000
     \end{array}\right] \nonumber \\
 \Lambda^{(1T)}_{H} &=& \left[ \begin{array}{r r r r}
       1.000 \hspace{0.3cm}&0.333 \hspace{0.3cm}&0.333 \hspace{0.3cm}&0.333         \\[0.3em]
       0.333 \hspace{0.3cm}&0.111 \hspace{0.3cm}&0.778 \hspace{0.3cm}&0.778         \\[0.3em]
       0.333 \hspace{0.3cm}&0.778 \hspace{0.3cm}&0.111 \hspace{0.3cm}&0.778        \\[0.3em]
       0.333 \hspace{0.3cm}&0.778 \hspace{0.3cm}&0.778 \hspace{0.3cm}&0.111         
     \end{array}\right] \nonumber \\
 \Lambda^{(1T)}_{Q} &=& \left[ \begin{array}{r r r r}
       1.000 \hspace{0.3cm}&0.333 \hspace{0.3cm}&0.333 \hspace{0.3cm}&0.333         \\[0.3em]
       0.333 \hspace{0.3cm}&0.556 \hspace{0.3cm}&0.940 \hspace{0.3cm}&0.171         \\[0.3em]
       0.333 \hspace{0.3cm}&0.171 \hspace{0.3cm}&0.556 \hspace{0.3cm}&0.940         \\[0.3em]
       0.333 \hspace{0.3cm}&0.940 \hspace{0.3cm}&0.171 \hspace{0.3cm}&0.556
     \end{array}\right] \nonumber \\
 \Lambda^{(2E)}_{free} &=& \left[ \begin{array}{r r r r}
       1.000 \hspace{0.3cm}& 0.326 \hspace{0.3cm}&  0.333  \hspace{0.3cm}&   0.347 \\[0.3em]
       0.351  \hspace{0.3cm}& 1.006  \hspace{0.3cm}& 0.336 \hspace{0.3cm}&0.314 \\[0.3em]
       0.325  \hspace{0.3cm}& 0.332 \hspace{0.3cm}&1.003\hspace{0.3cm}& 0.349 \\[0.3em]
       0.341  \hspace{0.3cm}&  0.336 \hspace{0.3cm}&0.337 \hspace{0.3cm}&1.011 
     \end{array}\right] \nonumber 
      \end{eqnarray} 
      \begin{eqnarray}
\Lambda^{(2E)}_{H} &=& \left[ \begin{array}{r r r r}
1.000  \hspace{0.3cm}&  2.159 \hspace{0.3cm}&   0.407  \hspace{0.3cm}&  0.841         \\[0.3em]
    2.147  \hspace{0.3cm}&  0.422  \hspace{0.3cm}&  0.752   \hspace{0.3cm}& 1.128          \\[0.3em]
    0.368  \hspace{0.3cm}&  0.787 \hspace{0.3cm}&  1.210  \hspace{0.3cm}&  2.151            \\[0.3em]
    0.876  \hspace{0.3cm}&  1.109  \hspace{0.3cm}&  2.147  \hspace{0.3cm}&  0.368         \\[0.3em]
             \end{array}\right] \nonumber \\            
\Lambda^{(2E)}_{Q} &=& \left[ \begin{array}{r r r r}
       1.000 \hspace{0.3cm}&0.762 \hspace{0.3cm}&0.018  \hspace{0.3cm}& 0.976         \\[0.3em]
       1.032 \hspace{0.3cm}&0.828 \hspace{0.3cm}&0.884  \hspace{0.3cm}&0.016        \\[0.3em]
       0.674 \hspace{0.3cm}&0.049 \hspace{0.3cm}&1.056  \hspace{0.3cm}&0.990         \\[0.3em]
      0.038 \hspace{0.3cm}&1.110 \hspace{0.3cm}&0.788  \hspace{0.3cm}&0.800
     \end{array}\right] \nonumber \\
 \Lambda^{(2T)}_{free} &=& \left[ \begin{array}{r r r r}
       1.000 \hspace{0.3cm}&0.329 \hspace{0.3cm}&0.328 \hspace{0.3cm}&0.338        \\[0.3em]
       0.329 \hspace{0.3cm}&1.000 \hspace{0.3cm}&0.336 \hspace{0.3cm}&0.335         \\[0.3em]
       0.328 \hspace{0.3cm}&0.336 \hspace{0.3cm}&1.000 \hspace{0.3cm}&0.335         \\[0.3em]
       0.338 \hspace{0.3cm}&0.335 \hspace{0.3cm}&0.335 \hspace{0.3cm}&1.000
     \end{array}\right] \nonumber \\
 \Lambda^{(2T)}_{H} &=& \left[ \begin{array}{r r r r}
       1.000 \hspace{0.3cm}&   2.085 \hspace{0.3cm}&   0.380 \hspace{0.3cm}&   0.859      \\[0.3em]
    2.085  \hspace{0.3cm}&  0.400  \hspace{0.3cm}&  0.758  \hspace{0.3cm}&  1.076         \\[0.3em]
    0.380 \hspace{0.3cm}&   0.758 \hspace{0.3cm}&   1.153  \hspace{0.3cm}&  2.054             \\[0.3em]
    0.859   \hspace{0.3cm}&  1.076 \hspace{0.3cm}&   2.054  \hspace{0.3cm}&  0.329   
     \end{array}\right] \nonumber \\     
 \Lambda^{(2T)}_{Q} &=& \left[ \begin{array}{r r r r}
       1.000 \hspace{0.3cm}&0.757 \hspace{0.3cm}& 0.179          \hspace{0.3cm}&0.962         \\[0.3em]
        1.011  \hspace{0.3cm}&0.810  \hspace{0.3cm}&0.891       \hspace{0.3cm}& 0.118        \\[0.3em]
       0.671 \hspace{0.3cm}& 0.048 \hspace{0.3cm}&  1.048 \hspace{0.3cm}&0.973         \\[0.3em]
       0.042 \hspace{0.3cm}& 1.126  \hspace{0.3cm}&0.784  \hspace{0.3cm}& 0.788
     \end{array}\right] \nonumber \\
  \Lambda^{(3E)}_{free} &=& \left[ \begin{array}{r r r r}
      1.000   \hspace{0.3cm}&  0.344   \hspace{0.3cm}&   0.325   \hspace{0.3cm}&  0.329 \\[0.3em]
       0.337  \hspace{0.3cm}& 1.004  \hspace{0.3cm}& 0.341  \hspace{0.3cm}&0.331 \\[0.3em]
      0.335     \hspace{0.3cm}&   0.328  \hspace{0.3cm}&  1.005 \hspace{0.3cm}&  0.352 \\[0.3em]
        0.327  \hspace{0.3cm}&  0.348  \hspace{0.3cm}& 0.329  \hspace{0.3cm}& 1.007
     \end{array}\right] \nonumber \\
 \Lambda^{(3E)}_{H} &=& \left[ \begin{array}{r r r r}
1.000 \hspace{0.3cm}&   0.980 \hspace{0.3cm}&   2.960  \hspace{0.3cm}&  0.995          \\[0.3em]
    0.955  \hspace{0.3cm}&  0.965  \hspace{0.3cm}&  1.000  \hspace{0.3cm}&  2.886              \\[0.3em]
    2.965  \hspace{0.3cm}&  0.985 \hspace{0.3cm}&   1.000  \hspace{0.3cm}&  0.990       \\[0.3em]
    0.990  \hspace{0.3cm}&  2.911  \hspace{0.3cm}&  0.990 \hspace{0.3cm}&   0.965           
     \end{array}\right] \nonumber \\
 \Lambda^{(3E)}_{Q} &=& \left[ \begin{array}{r r r r}
1.000 \hspace{0.3cm}& 1.013  \hspace{0.3cm}&   1.000 \hspace{0.3cm}& 0.000           \\[0.3em]
0.000 \hspace{0.3cm}&   1.025   \hspace{0.3cm}&   1.011 \hspace{0.3cm}& 1.020            \\[0.3em]
0.998  \hspace{0.3cm}&  0.000 \hspace{0.3cm}&  1.007  \hspace{0.3cm}& 1.007          \\[0.3em]
1.002  \hspace{0.3cm}&   1.007  \hspace{0.3cm}&  0.000 \hspace{0.3cm}&  1.009
\end{array}\right] \nonumber \\
 \Lambda^{(3T)}_{free} &=& \left[ \begin{array}{r r r r}
       1.000 \hspace{0.3cm}&0.330 \hspace{0.3cm}&0.336 \hspace{0.3cm}&0.330         \\[0.3em]
       0.330 \hspace{0.3cm}&1.000 \hspace{0.3cm}&0.337 \hspace{0.3cm}&0.336         \\[0.3em]
       0.336 \hspace{0.3cm}&0.337 \hspace{0.3cm}&1.000 \hspace{0.3cm}&0.332         \\[0.3em]
       0.330 \hspace{0.3cm}&0.336 \hspace{0.3cm}&0.332 \hspace{0.3cm}&1.000
     \end{array}\right] \nonumber \\
 \Lambda^{(3T)}_{H} &=& \left[ \begin{array}{r r r r}
      1.000  \hspace{0.3cm}&  0.995  \hspace{0.3cm}&  2.974 \hspace{0.3cm}&   0.994         \\[0.3em]
    0.994 \hspace{0.3cm}&   0.999  \hspace{0.3cm}&  0.974  \hspace{0.3cm}&  2.974             \\[0.3em]
    2.974 \hspace{0.3cm}&   0.974 \hspace{0.3cm}&   1.000 \hspace{0.3cm}&   0.987                \\[0.3em]
    0.994  \hspace{0.3cm}&  2.974  \hspace{0.3cm}&  0.987  \hspace{0.3cm}&  0.999
     \end{array}\right] \nonumber 
     \end{eqnarray}
     \begin{eqnarray}
 \Lambda^{(3T)}_{Q} &=& \left[ \begin{array}{r r r r}
 1.000  \hspace{0.3cm}&  0.994  \hspace{0.3cm}&  1.003  \hspace{0.3cm}&       0.000         \\[0.3em]
 0.000 \hspace{0.3cm}&   1.000  \hspace{0.3cm}&  0.994   \hspace{0.3cm}& 1.007             \\[0.3em]
 0.997   \hspace{0.3cm}&      0.000  \hspace{0.3cm}&  1.000  \hspace{0.3cm}&  0.994            \\[0.3em]
 0.994 \hspace{0.3cm}&   0.993    \hspace{0.3cm}&     0.000   \hspace{0.3cm}& 1.000
     \end{array}\right] \nonumber \\
 M^{(E)}_{free} &=& \left[ \begin{array}{r r r r}
        1.000  \hspace{0.3cm}&   -0.006  \hspace{0.3cm}&  0.004 \hspace{0.4cm}&  0.003 \\[0.3em]
      0.000  \hspace{0.3cm}&  0.994 \hspace{0.3cm}&  0.012  \hspace{0.4cm}& -0.006 \\[0.3em]
       0.001  \hspace{0.3cm}& -0.020 \hspace{0.3cm}&   0.981 \hspace{0.4cm}& 0.017 \\[0.3em]
       0.000 \hspace{0.3cm}& 0.001 \hspace{0.3cm}&  -0.008  \hspace{0.4cm}& 1.006
     \end{array}\right] \nonumber \\
 M^{(E)}_{H} &=& \left[ \begin{array}{r r r r}
        1.000  \hspace{0.2cm}& -0.008 \hspace{0.3cm}&  -0.007 \hspace{0.3cm}&  -0.002     \\[0.3em]
   -0.004 \hspace{0.2cm}&    1.010  \hspace{0.3cm}&   0.008 \hspace{0.3cm}&  -0.008    \\[0.3em]
   -0.003  \hspace{0.2cm}& -0.003 \hspace{0.3cm}&   -0.990  \hspace{0.3cm}&  -0.016    \\[0.3em]
   -0.002  \hspace{0.2cm}& -0.002  \hspace{0.3cm}&    0.001 \hspace{0.3cm}&  -1.008
     
     \end{array}\right] \nonumber \\
 M^{(E)}_{Q} &=& \left[ \begin{array}{r r r r}
        1.000         \hspace{0.2cm}& -0.007  \hspace{0.3cm}&  -0.002 \hspace{0.3cm}&    0.006    \\[0.3em]
    -0.008        \hspace{0.2cm}&  1.006   \hspace{0.3cm}&   0.012 \hspace{0.3cm}&   -0.013       \\[0.3em]
     -0.002            \hspace{0.2cm}& 0.007 \hspace{0.3cm}&    -0.001  \hspace{0.3cm}&  0.995     \\[0.3em]
   -0.006        \hspace{0.2cm}&   -0.002  \hspace{0.3cm}& -0.994  \hspace{0.3cm}&   -0.001
     \end{array}\right] \nonumber 
\end{eqnarray}
\begin{eqnarray}
 M^{(T)}_{free} &=& \left[ \begin{array}{r r r r}
       1 \hspace{0.3cm}&0 \hspace{0.35cm}&0 \hspace{0.3cm}&0         \\[0.3em]
       0 \hspace{0.3cm}&1 \hspace{0.35cm}&0 \hspace{0.3cm}&0         \\[0.3em]
       0 \hspace{0.3cm}&0 \hspace{0.35cm}&1 \hspace{0.3cm}&0         \\[0.3em]
       0 \hspace{0.3cm}&0 \hspace{0.35cm}&0 \hspace{0.3cm}&1
     \end{array}\right] \nonumber \\
 M^{(T)}_{H} &=& \left[ \begin{array}{r r r r}
       1 \hspace{0.3cm}&0 \hspace{0.1cm}&0 \hspace{0.1cm}&0         \\[0.3em]
       0 \hspace{0.3cm}&1 \hspace{0.1cm}&0 \hspace{0.1cm}&0         \\[0.3em]
       0 \hspace{0.3cm}&0 \hspace{0.1cm}&-1 \hspace{0.1cm}&0        \\[0.3em]
       0 \hspace{0.3cm}&0 \hspace{0.1cm}&0 \hspace{0.1cm}&-1         
     \end{array}\right] \nonumber \\
 M^{(T)}_{Q} &=& \left[ \begin{array}{r r r r}
       1 \hspace{0.3cm}&0 \hspace{0.15cm}&0 \hspace{0.3cm}&0         \\[0.3em]
       0 \hspace{0.3cm}&1 \hspace{0.15cm}&0 \hspace{0.3cm}&0         \\[0.3em]
       0 \hspace{0.3cm}&0 \hspace{0.15cm}&0 \hspace{0.3cm}&1         \\[0.3em]
       0 \hspace{0.3cm}&0 \hspace{0.15cm}&-1 \hspace{0.3cm}&0
     \end{array}\right] \nonumber 
\end{eqnarray}           

 We have verified that the experimental Mueller matrices are physically realizable or not with calculations of N-matrix discussed in \cite{nmatrix}. The N-matrices for three experimental Mueller matrices are given by 
    \newpage                    
\begin{widetext}
\begin{eqnarray} 
N_{free} &=& \left[ \begin{array}{r r r r}
      1.000        \hspace{0.2cm}&   0.008 - 0.002i \hspace{0.2cm}&    -0.010 - 0.000i \hspace{0.2cm}&   0.999 + 0.013i \\[0.3em]
    0.008 + 0.002i  \hspace{0.2cm}&    0.006     \hspace{0.2cm}&     -0.013 - 0.004i \hspace{0.2cm}&   0.011 + 0.001i   \\[0.3em]
  -0.010 + 0.000i \hspace{0.2cm}& -0.013 + 0.004i  \hspace{0.2cm}&    0.000     \hspace{0.2cm}&     -0.004 + 0.005i  \\[0.3em]
    0.999 - 0.013i \hspace{0.2cm}&  0.011 - 0.001i \hspace{0.2cm}&   -0.004 - 0.005i \hspace{0.2cm}&    1.006       
     \end{array}\right] \nonumber \\
 N_{H} &=& \left[ \begin{array}{r r r r}
      1.000      \hspace{0.2cm}&    0.000 - 0.005i \hspace{0.2cm}&  -0.003 + 0.002i \hspace{0.2cm}& -1.000 - 0.009i   \\[0.3em]
   0.000 + 0.005i \hspace{0.2cm}& -0.003      \hspace{0.2cm} &      0.009 + 0.008i \hspace{0.2cm}&  -0.000 + 0.000i    \\[0.3em]
  -0.003 - 0.002i \hspace{0.2cm}&  0.009 - 0.008i \hspace{0.2cm}&  -0.007      \hspace{0.2cm}&    -0.008 + 0.003i    \\[0.3em]
  -1.000 + 0.009i  \hspace{0.2cm}& -0.000 - 0.000i \hspace{0.2cm}&   -0.008 - 0.003i \hspace{0.2cm}& 1.011          
\end{array}\right] \nonumber \\     
 N_{Q} &=& \left[ \begin{array}{r r r r}
       1.000   \hspace{0.2cm}&      0.005 - 0.003i  \hspace{0.2cm}&  0.003 + 0.004i \hspace{0.2cm}&    -0.001 + 0.999i  \\[0.3em]
  0.005 + 0.003i   \hspace{0.2cm}& -0.004 \hspace{0.2cm}&      0.000 - 0.000i  \hspace{0.2cm}&    -0.004 + 0.002i    \\[0.3em]
  0.003 - 0.004i   \hspace{0.2cm}&   0.000 + 0.000i  \hspace{0.2cm}&  -0.002  \hspace{0.2cm}&     -0.007 + 0.009i      \\[0.3em]
   -0.001 - 0.999i     \hspace{0.2cm}&   -0.004 - 0.002i  \hspace{0.2cm}&   -0.007 - 0.009i  \hspace{0.2cm}&   1.015
     \end{array}\right] \nonumber 
\end{eqnarray}
\end{widetext}   
                     
The eigenvalues of the N-matrices are ($-$0.0117, 0.0036, 0.0176, \textbf{2.0024}), ($-$0.0189, 0.0039, 0.0111, \textbf{2.0055}) and ($-$0.0079, $-$0.0033, 0.0136,  \textbf{2.0071}) for free space, half wave plate and quarter wave plate respectively. From these values, it is clear that all eigenvalues are real and only one eigenvalue is non-zero. It shows that N-matrices are hermitian and the Mueller matrices represent the real physical systems. Apart from this, we have also verified the following inequalities 
 \begin{equation} \label{dpl}
 N^{2} = tr(N).N,
 \end{equation}
\begin{equation}
tr(N) = 2m_{00}  \hspace{0.3cm} \rm{and} \hspace{0.3cm} tr(MM^T)  \le 4m_{00}^{2}. 
 \end{equation}
Eq. (\ref{dpl}) proves that the chosen optical elements are non-depolarizing elements. The last equation is a constrain on degree of polarization for a physical system, that is also satisfied in our case \cite{depo}. We have calculated the retardance for QWP and HWP by decomposing their Mueller matrices \cite{chipman}. The retardations of HWP and QWP are 0.997$\pi$, 0.501$\pi$ respectively. These errors are with in the quoted tolerances by the manufacturers. The experimental conditions are same for all measurements i.e. there is no necessity of inserting or removing optical components during the experimental observations, which minimizes the errors in Mueller matrix elements.  

\section{Error Analysis}

  The main sources of errors in this set up are the deviations in fast axes of wave plates from vertical axis and errors in their retardance. Along with these, fluctuations in the intensity of light and the measurement by the power meter also contribute to errors in Mueller matrix of the element. We have neglected the error due to tilt of the wave plates which has been taken care in the alignment of the set up. The error analysis of the total system is done by the procedure followed in \cite{error, error1, error2}. The whole experimental set up can be realized with the Jone's matrices and written as \cite{gold}
  
\begin{eqnarray} \label{eq:full_exp}
\Lambda_{ij} &=& {I}\left[~{J_{P}}(\theta_{P2})\centerdot{J_{R}}(\phi_{RQ},\theta_{R6j})\centerdot\right.\\ \nonumber
     &~& {J_{R}}(\phi_{RH},\theta_{R5j})\centerdot{J_{R}}(\phi_{RQ},\theta_{R4j})\centerdot O \centerdot\\ \nonumber
     &~& {J_{R}}(\phi_{RQ},\theta_{R3i})\centerdot{J_{R}}(\phi_{RH},\theta_{R2i})\centerdot\\ \nonumber
     &~& \left.{J_{R}}(\phi_{RQ},\theta_{R1i})\centerdot{J_{P}}(\theta_{P1})\centerdot{L}\right]
\end{eqnarray}
where $ I $ denotes the intensity recorded by the detector, ${J_{P}}(\theta)$, ${J_{R}}(\phi,\theta)$ denote Jone's matrices for the polarizer rotated by an angle $\theta$ and the retarders having retardation $\phi$ rotated by an angle $\theta$ respectively, $RH$ and $RQ$ denote the retardation offered by HWP and QWP respectively, $P1$ and $P2$ are the angles of rotation of polarizers, $O$ is the object for which the Mueller matrix is determined  and $ L $ denotes the laser. $\Lambda$ denotes the power shown by the power-meter. In Jones vector study, $ L $ representing a vertically polarized light with a $2\times1$ column vector and all other optics are $2\times2$ matrices. The full quantity inside the function $ I$ is a $2\times1$ matrix. The power meter values provide the elements of the projection matrix. From the different manufactures of the optics and rotation stages, we obtained the tolerances or mean error introduced by the individual elements. Considering these errors to be uncorrelated, we can write the variance introduced in $\Lambda$ due to all the optical elements is \cite{error1}
\begin{equation}\label{eq:sigma}
\sigma_{\Lambda_{ij}}=\sqrt{ \sum_{{\rm all}~x} \left(\frac{\partial f}{\partial x}\right)^2\sigma_x^2    }
\end{equation}
where $x$'s are the variables from Eq. (\ref{eq:full_exp}). Tolerances in optical elements and their rotation angles are
\begin{eqnarray} \label{eq:errors}
\sigma\theta_{P1}&=&0.03^{\rm o} \nonumber \\ 
\sigma\theta_{P2}&=&0.03^{\rm o} \nonumber \\
\sigma\phi_{RH}&=&1.26^{\rm o} \nonumber \\
\sigma\phi_{RQ}&=&1.26^{\rm o} \nonumber\\
\sigma\theta_{Ri}&=&0.04^{\rm o},\quad {\rm for}~i=1~{\rm to}~6.
\end{eqnarray}
Using eqs. \ref{eq:full_exp}-\ref{eq:errors}, we can evaluate the errors corresponding to all the elements of $\Lambda$ matrix. After obtaining the error in projection elements, we have introduced the 1\% intensity variation due to laser and 0.2\% due to the detector. Now we do the same error propagation analysis with 16 variables present in projection matrix to obtain the error in Mueller matrix elements as
\begin{eqnarray}
\label{eq:error1}
 \sigma_{M} =\left[ \begin{array}{r r r r}
       0.007 &  0.012 &  0.012 & 0.013 \\[0.3em]
       0.012 &  0.021 &  0.021 & 0.022 \\[0.3em]
       0.012 &  0.021 &  0.019 & 0.020 \\[0.3em]
       0.013 &  0.022 &  0.020 & 0.026
     \end{array}\right].
\end{eqnarray}
Experimentally observed errors are close to the theoretically calculated errors in Mueller matrix. In our experiment, the maximum error obtained in individual elements while calculating Mueller matrices are $\pm$0.020, $\pm$0.016, $\pm$0.013 for free space, half wave plate and quarter wave plate respectively. These errors are within the limits as calculated in eq. \ref{eq:error1}.
  
Now we compare our results with other methods based on polarimetry. The maximum error in the Mueller matrix determined by Dev et al.\cite{comp} with simple polarimetry is $\pm$0.0985.  Goldstein et al. \cite{com} experimentally determined Mueller matrix with the maximum error of $\pm$0.034 by using dual rotating retarder method. Bueno \cite{bueno} obtained Mueller matrix with the maximum error of $\pm$0.014 by using the liquid crystal variable retarders whose retardance is changed by applied voltage. Baba et al. determined Mueller matrix by imaging polarimetry that uses variable rotators and retarders \cite{MMIP}.  They have obtained the maximum error $\pm$0.0353 by taking 16 images and $\pm$0.0138 while taking 36 images. We have taken the Mueller matrix of \textit{free space} for comparing the error with other methods. The present method uses simple wave plates and takes just 16 measurements.    \\

\section{Conclusions}

We have discussed a new and simple technique to determine the Mueller matrix of an optical element by using two SM gadgets and it's efficacy has been verified for three known optical elements i.e. free space, half wave plate and quarter wave plate. One may use this gadget in quantum process tomography \cite{QPT} and polarization state tomography \cite{QCD} which finds diverse applications in quantum information.


\begin{thebibliography}{99}

\bibitem{chipman} S. Y. Lu, and R. A. Chipman, ``Interpretation of Mueller matrices
based on polar decomposition'', J. Opt. Soc. Am. A \textbf{13}, 1106 (1996).

\bibitem{nghosh} S. Manhas, M. K. Swami, P. Buddhiwant, N. Ghosh, P. K. Gupta, and K. Singh, ``Mueller matrix approach for determination of optical rotation in chiral turbid media in backscattering geometry'', Optics Express \textbf{14}, 190 (2006).

\bibitem{OCT} W. C. Kuo, N. K. Chou, C. Chou, C. M. Lai, H. J. Huang, S. S Wang, and J. J. Shyu, ``Polarization-sensitive optical coherence tomography for imaging human atherosclerosis'', Applied  Optics \textbf{36}, 2520 (2007).

\bibitem{micro-elec} T. Novikova, A. De Martino, S. B. Hatit, and B. Drévillon, ``Application of Mueller polarimetry in conical diffraction for critical dimension measurements in microelectronics'', Applied Optics \textbf{45}, 3688 (2006).

\bibitem{ocean} K. J. Voss, and E. S. Fry, ``Measurement of the Mueller matrix for ocean water'', Applied Optics \textbf{23}, 4427 (1984).

\bibitem{astro} J. Tinbergen, \textit{Astronomical Polarimetry}, Cambridge University Press (1996).

\bibitem{rpsir} V. K. Jaiswal, R. P. Singh, and R. Simon, ``Producing optical vortices through forked holographic grating: study of polarization'', Journal of Modern optics \textbf{57}, 2031 (2010).

\bibitem{SLM} I. Moreno, A. Lizana, J. Campos, A. Márquez, C. Iemmi, and M. J. Yzuel, ``Combined Mueller and Jones matrix method for the evaluation of the complex modulation in a liquid-crystal-on-silicon display'', Optics Letters \textbf{33}, 627 (2008).

\bibitem{tomography} C. K. Hitzenberger, E. Götzinger, M. Sticker, M. Pircher, and
A. F. Fercher, ``Measurement and imaging of birefringence and optic axis orientation by phase resolved polarization sensitive optical coherence tomography'', Optics Express \textbf{9}, 780 (2001).

\bibitem{Simon} R. Simon, and N. Mukunda, ``Minimal three-component SU(2) gadget for polarization optics'', Physics Letters A \textbf {143}, 165 (1990).

\bibitem{bagini} V. Bagini, R. Borghi, F. Gori, M. Santarsiero, F. Frezza, G. Schettini, and G. S. Spagnolo, ``The Simon–Mukunda polarization gadget'', Euro Journal of Physics \textbf{17}, 279 (1996).

\bibitem{Neeti} B. Neethi Simon, C. M. Chandrashekar, and S. Simon, ``Hamilton's turns as visual tool-kit for designing of single qubit unitary gates'', Physical Review A \textbf{85}, 022323 (2012).

\bibitem{lidar} M. Hayman, and J. P. Thayer, ``Lidar polarization measurements of PMCs'', Journal of Atmospheric and Solar-Terrestrial Physics \textbf{73}, 2110 (2011).

\bibitem{sm} U. Schilling, J. V. Zanthier, and G. S. Agarwal, ``Measuring arbitrary-order coherences: Tomography of single-mode multiphoton polarization-entangled states'', Physical Review A \textbf{81}, 013826 (2010).

\bibitem{pancha} J. C. Loredo, O. Ortíz, R. Weingärtner, and F. De Zela, ``Measurement of Pancharatnam phase by robust interferometric and polarimetric methods'', Physical Review A \textbf{80}, 012113 (2009).

\bibitem{layden} D. Layden, M. F. G. Wood, and I. A. Vitkin, ``Optimum selection of input polarization states in determining the sample Mueller matrix: a dual photoelastic polarimeter approach'', Optics Express \textbf{20}, 20466 (2012).

\bibitem{mmip} W. S. Bicke, and W. M Bailey, ``Stoke's vectors, Mueller matrices and polarized light scattering'', Am. J. Phys.  \textbf{53}, 468 (1985).

\bibitem{azaam} R. M. A. Azzam, ``Photopolarimetric measurement of the Mueller matrix
by Fourier analysis of a single detected signal'', Optics Letters \textbf{2}, 148 (1978).


\bibitem{gold} D. Goldstein, \textit{Polarized light}, Marcel Dekker publishers (2003). 

\bibitem{nmatrix}  R. Simon, ``The connection between Mueller and Jone's matrices of polarization optics'', Optics Communications \textbf{42}, 293 (1982).

\bibitem{depo} E. S. Fry, and G. W. Kattawar, ``Relationships between elements of the Stokes matrix'', Applied Optics \textbf{20}, 2811 (1981).


\bibitem{error} H. H. Ku, ``Notes on the Use of Propagation of Error Formulas'', Journal of Research of the National Bureau of Standards - C. Engineering and Instrumentation, \textbf{70C}, 4 (1966).

\bibitem{error1} A. C. Melissinos, Experiments in Modern Physics, Academic press, New York (1966), Sec 10.4, pp 467-479.

\bibitem{error2} D. F. V. James, P. G. Kwait, W. J. Munro, and A. G. White, ``Measurement of qubits'', Phys. Rev. A \textbf{64}, 052312 (2001).

\bibitem{com} D. H. Goldstein, ``Mueller matrix dual-rotating retarder polarimeter'', Applied Optics \textbf{31}, 6676 (1992).

\bibitem{comp} K. Dev, and A. Asundi, ``Mueller Stokes polarimetric characterization of transmissive liquid crystal spatial light modulator'', Optics and Lasers in Engineering \textbf{50}, 599  (2012).

\bibitem{bueno} J. M. Bueno, ``Polarimetry using liquid-crystal variable retarders: theory and calibration'', Journal of Optics A: Pure and Applied Optics, \textbf{2}, 216 (2000).

\bibitem{MMIP} J. S. Baba, J. R. Chung, A. H. DeLaughter, B. D. Cameron, and G. L. Cote, ``Development and calibration of an automated Mueller matrix polarization imaging system'', Journal of Biomedical Optics \textbf{7(3)}, 341 (2002).

\bibitem{QCD} A. Ling, K. P. Soh, A. L. Linares, and C. Kurtsiefer, ``Experimental polarization state tomography using optimal polarimeters'', Physical Review A \textbf{74}, 022309  (2006).

\bibitem{QPT} M. W. Mitchell, C. W. Ellenor, S. Schneider, and A. M. Steinberg, ``Diagnosis, Prescription, and Prognosis of a Bell-State Filter by Quantum Process Tomography'', Physical Review Letters \textbf{91}, 120402 (2003).
\end{thebibliography}
\end{document}